\documentclass{jfm}

\usepackage{graphicx}
\usepackage{newtxtext}
\usepackage{newtxmath}
\usepackage{natbib}
\usepackage{hyperref}
\usepackage{natbib}
\usepackage{amsmath,mathtools}
\hypersetup{
    colorlinks = true,
    urlcolor   = blue,
    citecolor  = black,
}
\usepackage{color}
\usepackage{soul}

\usepackage{tikz}
\usetikzlibrary{arrows,arrows.meta}
\usepackage{bm}

\newcommand{\RomanNumeralCaps}[1]
\linenumbers

\newcommand\barL[1]{\bar{#1}^{\scriptscriptstyle \mathrm{L}}}

\newcommand{\bphi}{\boldsymbol{\varphi}}
\newcommand{\ba}{\boldsymbol{a}}
\newcommand{\bu}{\boldsymbol{u}}

\newcommand{\bx}{\boldsymbol{x}}
\newcommand{\bX}{\boldsymbol{X}}
\newcommand{\bY}{\boldsymbol{Y}}
\let\Xi\varXi
\newcommand{\bxi}{\boldsymbol{\xi}}
\newcommand{\bXi}{\boldsymbol{\Xi}}

\def\beq{\begin{equation}}
\def\eeq{\end{equation}}
\def\d{\mathrm{d}}
\def\lie{\mathcal{L}}

\def\e{\mathrm{e}}

\makeatletter
\newcommand{\oset}[3][0ex]{%
  \mathrel{\mathop{#3}\limits^{
    \vbox to#1{\kern-1\ex@
    \hbox{$\scriptstyle#2$}\vss}}}}
\makeatother

\newcommand\BARL[1]{{\oset{-\phantom{\mkern-5.5mu\rhook}}{#1}}{\vphantom{\bar{#1}}}^{\scriptscriptstyle \mathrm{L}}}
\newcommand\tilL[1]{\tilde{#1}^{\scriptscriptstyle \mathrm{L}}}
\newcommand\TILDL[1]{{\tilde{#1}\mkern1.5mu}{\vphantom{\bar{#1}}}^{\scriptscriptstyle \mathrm{L}}}

\newcommand\BAR[1]{\bar{#1}}
\newcommand{\barLbphi}{\BARL{\bphi}}
\newcommand\HAKrev[1]{{#1}}
\newcommand\JVrev[1]{{#1}}
\newcommand\revv[1]{{#1}}

\title{Computing Lagrangian means}

\author{Hossein A. Kafiabad\aff{1}
  \corresp{\email{h.kafiabad@ed.ac.uk}} \and
  Jacques Vanneste\aff{1}}

\affiliation{\aff{1}School of Mathematics and Maxwell Institute for Mathematical Sciences, University of Edinburgh, EH9 3FD, UK}

\begin{document}
\maketitle

\begin{abstract}
Lagrangian averaging plays an important role in the analysis of wave--mean-flow interactions and other multiscale fluid phenomena. The numerical computation of Lagrangian means, e.g.\ from simulation data, is however challenging. Typical implementations require tracking a large number of particles to construct Lagrangian time series which are then averaged using a low-pass filter. This has drawbacks that include large memory demands, particle clustering and complications of parallelisation. 
We develop a novel approach in which the Lagrangian means of various fields (including particle positions) are computed by solving partial differential equations (PDEs) that are integrated over successive averaging time intervals. We propose two strategies, distinguished by their spatial independent  variables. The first, which generalises the algorithm of \citeauthor{GBLApaper} (2022, \textit{J. Fluid Mech.} \textbf{940}, A2), uses end-of-interval particle positions; the second directly uses the Lagrangian mean positions. The PDEs can be discretised in a variety of ways, e.g.\ using the same discretisation as that employed for the governing dynamical equations, and solved on-the-fly to minimise the memory footprint. 
We illustrate the new approach with a pseudospectral implementation for the rotating shallow-water model. \JVrev{Two applications to flows that combine vortical turbulence and  Poincar\'e waves demonstrate the superiority of Lagrangian averaging over Eulerian averaging for wave--vortex separation.}

%
%


\end{abstract}

%
%

\section{Introduction}
\label{sec:intro}

Time averaging is a basic yet essential tool in fluid dynamics because of the ubiquity of phenomena involving multiple time scales. Atmospheric and oceanic flows, for instance, can often be decomposed usefully into fast and slow components \citep[e.g.][]{vallis2017atmospheric,vanneste2013balance}. Time averaging is then used to interpret numerical-simulation and observational  data and to assimilate the latter in ocean and weather models. \JVrev{More broadly, it can be argued that all numerical simulations of time-dependent flows rely implicitly on time averaging, since they do not resolve phenomena on time scales shorter than the time step.} 
%
%
%
%

Time averaging can be performed in different ways. The most straightforward way is to average the time series of flow variables at fixed spatial positions to obtain the so-called \textit{Eulerian mean}. An alternative is to average flow variables along particle trajectories, that is, at fixed particle labels instead of fixed positions, to obtain the \textit{Lagrangian mean}. There are several practical and conceptual reasons to view Lagrangian averaging as superior to Eulerian averaging.

\begin{figure}
\centering
\includegraphics[width=\linewidth]{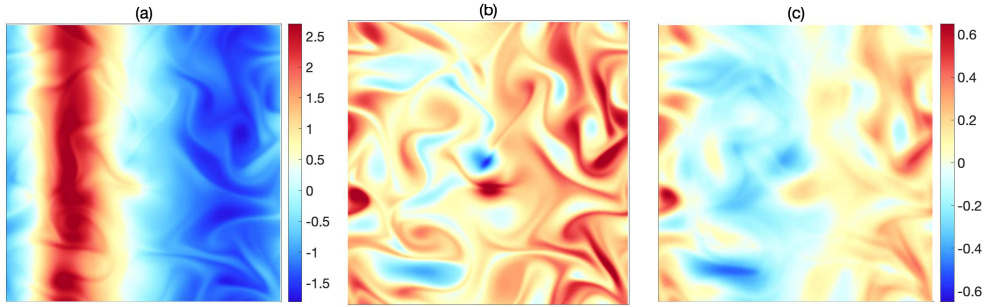}
\caption{Vorticity field and its Lagrangian and Eulerian means in the shallow-water simulation of \S\ref{sec:SW_example1}: (a) instantaneous vorticity, (b) Lagrangian mean vorticity, and (c) Eulerian mean vorticity. Both mean fields share the same averaging period corresponding to 3.6 wave periods. Panels (b) and (c) share the same colourbar shown to the right of panel (c).}
\label{fig:vorticities}
\end{figure}

From a practical viewpoint, the advection of fast motion (often consisting of waves) by slowly evolving flow and vice versa adversely affect their Eulerian decomposition. For instance, a strong, slowly evolving flow `Doppler shifts' the frequency of fast waves. This can lead to wave frequencies observed at fixed spatial locations that are much smaller than the intrinsic frequency. It therefore obscures the time-scale separation between waves and flow. Lagrangian averaging resolves this issue, as  demonstrated in several studies \citep{nagai2015spontaneous,shakespeare2017spontaneous,bachman2020particle,shakespeare2021new,jones2022separating}. Conversely, the advection of a slow background flow by waves leads to flow features that are blurred by Eulerian averaging but not by Lagrangian averaging. 
We illustrate this in figure \ref{fig:vorticities} showing the vorticity field in a simulation of a turbulent rotating shallow-water flow
interacting with a mode-1 Poincar\'e wave. The instantaneous vorticity in panel (a) is dominated by the high-amplitude wave. Both the Lagrangian and Eulerian means (panels (b) and (c) respectively) filter out the wave, but the Eulerian mean blurs out 
fine vorticity structures which are well resolved by the
Lagrangian mean (details of this simulation are presented in \S \ref{sec:SW_example1}).

From a conceptual viewpoint, Lagrangian averaging provides a powerful tool to study wave--mean-flow interactions. This is because the material conservation of key fields (scalar concentrations, vorticity vector, circulation, potential vorticity) is naturally inherited by the corresponding Lagrangian mean fields. As a result, the Lagrangian mean  of the dynamical equations is often simpler and more meaningful than the Eulerian mean  \citep{sowa72,bretherton1971general,andrews1978exact,buhler2009waves,gilbert2018geometric}. Relatedly, the Lagrangian mean  emerges naturally in the asymptotic derivation of wave--averaged models \citep{grimshaw1975nonlinear,wagner2015available}. A striking example of the dynamical relevance of the Lagrangian mean is provided by the observation that  geostrophic balance, the dominant balance in rapidly rotating flows, continues to hold in the presence of strong waves provided it is formulated in terms of Lagrangian mean velocity and pressure instead of their instantaneous or Eulerian mean values \citep{moore1970mass,buhler1998non-dissipative,kafiabad2021wave,GBLApaper}.

%

Despite its advantages, Lagrangian averaging is not widely used as a practical tool, mainly because Lagrangian means are difficult to compute numerically. Most numerical models are intrinsically Eulerian and provide the fields of interest at fixed spatial locations, typically grid points. The standard approach for the computation of Lagrangian means is then to seed a large number of passive particles in the flow, track them (forward or backward in time) using interpolated velocities, and apply time averaging to the resulting Lagrangian time series \citep[e.g.][]{nagai2015spontaneous,shakespeare2017spontaneous,shakespeare2018life,shakespeare2019momentum,shakespeare2021new,bachman2020particle,jones2022separating}. This has a high computational cost, requires a large memory allocation, suffers from possible particle clustering and, as discussed in \cite{GBLApaper}, is difficult to parallelise efficiently (see \citet{shakespeare2021new} for a parallel implementation).


To circumvent the difficulties of particle tracking, \citet{GBLApaper} developed a grid-based method that computes the Lagrangian mean directly on an Eulerian grid, building the mean through a time step iteration carried out over  successive averaging intervals. By eliminating the need to compute explicit particle trajectories, the method reduces memory demands and simplifies integration into  parallelised numerical models. The present paper starts with the recognition that the algorithm of \citet{GBLApaper} is a particular discretisation of a PDE governing the evolution of what we term \textit{partial Lagrangian mean}, that is, the mean carried out  only up to some intermediate time in the averaging interval. We formulate this PDE using the position of particles at the intermediate time as independent spatial variable,  as in \citet{GBLApaper}. The (total) Lagrangian mean is then obtained by taking the intermediate time to be the end of the averaging interval. 

In this form, the Lagrangian mean does not match \citeauthor{andrews1978exact}'s (\citeyear{andrews1978exact}) definition of the generalised Lagrangian mean (GLM): this requires the mean fields to be expressed as functions of the mean position of fluid particles. To achieve this, it is necessary to relate the mean positions of particles to their positions at the end of averaging interval, and to carry out a remapping of the Lagrangian mean fields. This constitutes our strategy 1 for the computation of generalised  Lagrangian means. We show that the algorithm of \citet{GBLApaper} amounts to a semi-Lagrangian discretisation of the PDEs of strategy 1. We propose an alternative strategy, strategy 2, which formulates PDEs directly for the partial Lagrangian means using the mean position as independent spatial variable. The PDEs involved in both strategies can be solved by broad classes of numerical methods: finite differences, finite volumes, finite elements or spectral methods. We illustrate this with a pseudospectral Fourier implementation for a shallow-water flow in a doubly periodic domain. 

The paper is structured as follows.  We introduce notation and define the Lagrangian means in \S \ref{sec:formulation}. We derive the PDEs of the two strategies in \S\ref{sec:PDEs}. We discuss their numerical implementation and present \JVrev{two} application to a shallow-water simulation in \S\ref{sec:num_impl}.  The choice of strategy, their advantages and costs are discussed in \S\ref{sec:discussion}. Technical aspects including the averaging of tensorial fields are relegated to appendices.

\section{Formulation}\label{sec:formulation}


We consider fluid motion in a two- or three-dimensional  Euclidean space. We denote the flow map by $\bphi$, with $\bphi(\ba,t) \in \mathbb{R}^2$ or $\mathbb{R}^3$ the position at time $t$ of a particle identified by its label $\ba$ (which can be taken as the position at $t=0$). The flow map and velocity field are related by
\beq
\partial_t \bphi(\ba,t) = \bu(\bphi(\ba,t),t).
\label{flowmap}
\eeq

Lagrangian averaging is averaging at fixed particle label $\ba$, in contrast with Eulerian averaging which fixes the spatial position. Both can involve different types of means: temporal, spatial or -- as often used in theoretical work -- ensemble mean.  Here we focus on a straigthforward time average, of the form
\beq
\BAR g(\tau) = \frac{1}{T} \int_{\tau}^{\tau + T} g(s) \, \d s
\label{ave}
\eeq
when applied to a function $g(t)$ that depends on time only. Eq.\ \eqref{ave} introduces the notation $\tau$ for the time at which the averaging is carried out and $T$ for the averaging period. Usually the middle of averaging interval $\tau + T/2$ is used as an argument for the mean function, but we prefer to adopt $\tau$ (the beginning of averaging interval) to simplify the notation in the upcoming derivations. A simple shift in time switches from one convention to the other. A weight function could be inserted in the integrand of \eqref{ave} to generalise the definition of the average; this would lead to minimal changes in what follows.

The Lagrangian mean trajectory associated with \eqref{ave} is represented by the Lagrangian mean map $\barLbphi$ defined by
\beq
\barLbphi(\ba,\tau) \coloneq \frac{1}{T} \int_{\tau}^{\tau + T}  \bphi(\ba,s) \, \d s. 
\label{barLbphi}
\eeq
Thus $\barLbphi(\ba,\tau)$ returns the mean position from $\tau$ to $\tau +T$ of the particle labelled by $\ba$. The definition \eqref{barLbphi} makes sense in $\mathbb{R}^n$, when $\barLbphi$ can be interpreted as a vector and averaged component-wise, but not on other manifolds where more complicated definitions are necessary \citep{gilbert2018geometric}. 
The (generalised) Lagrangian mean of a scalar function $f(\bx,t)$ is then defined by
\beq
\BARL f( \barLbphi(\ba,\tau), \tau) \coloneq \frac{1}{T} \int_{\tau}^{\tau + T} f(\bphi(\ba,s),s) \, \d s. 
\label{barLf}
\eeq
Hence $\BARL f(\bx,\tau)$ is the average of $f$ along the trajectory of the fluid parcel, regarded as a function of the Lagrangian mean position $\bx$ and time $\tau$. 

Our aim is the development of an efficient numerical method for the computation of $\BARL f$ that relies on solving PDEs, which can be discretised in a variety of ways, rather than on tracking ensembles of particle trajectories. We propose two strategies and derive the corresponding  PDEs in the next section.

\section{Two strategies} \label{sec:PDEs}

\begin{figure}
\begin{minipage}{.5\textwidth}
\centerline{(a)}
\begin{tikzpicture}[scale=0.8,
    thick,
    >=stealth',
    dot/.style = {
      draw,
      fill = white,
      circle,
      inner sep = 0pt,
      minimum size = 4pt
    }
  ]
  \coordinate (O) at (0,0);
  \coordinate (A) at (.5,5);
  \coordinate (B) at (6,.5);
  \coordinate (C) at (6,2.2);
  
  \draw[->] (-0.3,0) -- (6.9,0) coordinate[label = {below:$t$}] (xmax);
  \draw[->] (0,-0.3) -- (0,5.5) coordinate[label = {right:$f$}] (ymax);
  \draw[thick, blue] (A) .. controls (2.5, 1) and (4, 2)  ..
      (B);
  \draw[thick, purple] (A) .. controls (2.5, 2.4) and (4, 2.5)  ..
      (C) node[xshift=0em,yshift=.2em,anchor=south,align=center] {$\TILDL{f}(\bphi(\ba,\tau+T),\tau)$ \\ $= \BARL{f}(\barLbphi(\ba,\tau),\tau) $} ; 
   \draw[thick, blue, fill=blue!20] (A) circle (.8mm);
   \draw[thick, blue, fill=blue!20] (B) circle (.8mm);
   \draw[thick, purple, fill=purple!20] (C) circle (.8mm);
   \draw[dashed] (A) -- (A |- O) node[anchor=north,xshift=.0em] {$\tau$} ;
   \draw[dashed] (C) -- (C |- O) node[anchor=north] {$\tau + T$} ;
   \node[purple] at (3,3.8) {$\tilde{f}(\bm{\varphi}(\bm{a},t),t;\tau)$};
   \node[blue] at (1.6,2.) {$f(\bphi(\ba,t),t)$};
   
\end{tikzpicture}
\end{minipage}
\begin{minipage}{.5\textwidth}
\centerline{(b)}
\begin{tikzpicture}[scale=0.8,
    thick,
    >=stealth',
    dot/.style = {
      draw,
      fill = white,
      circle,
      inner sep = 0pt,
      minimum size = 4pt
    }
  ]
  \coordinate (O) at (0,0);
  \coordinate (A) at (.5,1);
  \coordinate (B) at (6,5);
  \coordinate (C) at (6,3);
  
  \draw[->] (-.3,0) -- (6.9,0) coordinate[label = {below:$t$}] (xmax);
  \draw[->] (0,-.3) -- (0,5.5) coordinate[label = {right:$\bm{x}$}] (ymax);

  \draw[thick, blue] (A) .. controls (2.5, 1) and (4, 2)  ..
      (B) node[xshift=.5em, anchor=south] {$\bphi(\ba,\tau + T)$};
  \draw[thick, purple] (A) .. controls (2.5, 1) and (4, 1)  ..
      (C) node[xshift=.5em, anchor=west] {$\barLbphi(\ba,\tau)$} ; 
   \draw[thick, blue, fill=blue!20] (A) circle (.8mm);
   \draw[thick, blue, fill=blue!20] (B) circle (.8mm);
   \draw[thick, purple, fill=purple!20] (C) circle (.8mm);
   \draw[dashed] (A) -- (A |- O) node[anchor=north,xshift=.0em] {$\tau$} ;
   \draw[dashed] (B) -- (B |- O) node[anchor=north] {$\tau + T$} ;
   \node[blue] at (3.3,2.7) {$\bphi(\bm{a},t)$};
   \node[purple] at (4.5,1.25) {$\bar{\bphi}(\ba,t;\tau)$};
   
\end{tikzpicture}
\end{minipage}
\caption{Lagrangian means and partial Lagrangian means in an averaging interval $(\tau, \tau + T)$: (a) means of a function $f$ evaluated along the trajectory $\bphi(\ba,t)$ of a fluid parcel labelled by $\ba$, and (b) means of the position, shown here along a single coordinate axis.}
\label{fig:oned}
\end{figure}
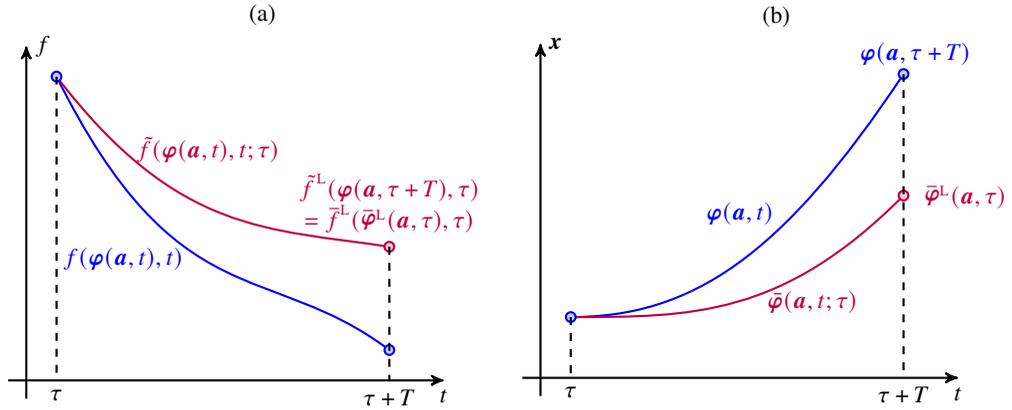

\JVrev{Following \citet{GBLApaper}, we introduce another representation of the Lagrangian mean via tilde functions defined by
\begin{subequations}
\begin{align}
 \TILDL{f}(\bphi(\ba,\tau + T), \tau)  &\coloneq \frac{1}{T} \int_{\tau}^{\tau + T} f(\bphi(\ba,s),s) \, \d s \label{tildeLf} \\
 &=  \BARL f( \barLbphi(\ba,\tau), \tau). \label{tildebar} 
\end{align}
\end{subequations}
}
Comparing this definition with \eqref{barLf} shows that the overbar indicates a mean along a trajectory identified by the Lagrangian mean position, whereas the tilde indicates the same mean but with the trajectory identified by the actual position at $t=\tau + T$, that is, at the end of the averaging. 
This is illustrated in figure \ref{fig:oned}. We also introduce the \textit{partial mean} versions of \eqref{barLbphi}, \eqref{barLf}  and \eqref{tildebar}, namely
\begin{subequations}
\label{partial_means}
\begin{align}
 \BAR{\bphi}(\ba,t;\tau) &\coloneq \frac{1}{t-\tau} \int_{\tau}^{t}  \bphi(\ba,s) \, \d s,  \label{barphi} \\
\BAR f( \BAR{\bphi}(\ba,t; \tau), t; \tau) &\coloneq \frac{1}{t-\tau} \int_{\tau}^{t}  f(\bphi(\ba,s),s) \, \d s,  \label{fbar}  \\
\tilde f(\bphi(\ba,t),t;\tau) &\coloneq \frac{1}{t-\tau} \int_{\tau}^t f(\bphi(\ba,s),s) \, \d s, \label{ftilde}
\end{align} 
\end{subequations}
as in \citet{GBLApaper}; see figure \ref{fig:oned}. Clearly the partial means give the total means when evaluated at $t = \tau + T$.  
We emphasise that all means used in the paper are Lagrangian means and that we only indicate this explicitly by a superscript L for the total means, to distinguish them from the partial means which are undecorated. The counterpart of \eqref{tildebar} holds for the partial means:
\beq
	\tilde f(\bphi(\ba,t),t;\tau)  = \BAR f( \BAR{\bphi}(\ba,t; \tau), t; \tau) \label{ftilde2fbar}.
\eeq

Since time average quantities vary over time scales larger than the averaging period $T$, it is neither necessary nor desirable to compute Lagrangian means at each of the times at which the velocity $\bu$ and scalar field $f$ are known, typically discrete times separated by a small time step. Rather, we think of the averaging time $\tau$ as a slow variable and propose to compute the Lagrangian means only at $\tau = \tau_n = n T$ for $n=0,1,2,\cdots$. We can therefore carry out independent computations for each $t_n$, each involving only the fields for $t \in (\tau_n, \tau_n+T)$. We now focus on one such interval and, to lighten the notation, drop the parametric dependence on $\tau$ from the partial means in \eqref{partial_means}. For instance, we use $\BAR f( \BAR{\bphi}(\ba,t), t)$ instead of $\BAR f( \BAR{\bphi}(\ba,t; \tau), t; \tau)$ in the following derivations, keeping in mind the now implicit dependence of $\BAR f $ and $\BAR{\bphi}$ on $\tau$. 
\JVrev{A warning about notation might be useful: in what follows, we use the symbol $\bx$ as a generic dummy variable, without attributing it the specific meaning of either an actual or Lagrangian mean position. This meaning is determined by the function in which $\bx$ appears. Thus $\bx$ in $\tilde f(\bx,t)$ is interpreted as an actual position at time $t$, whereas in $\BAR f(\bx,t)$ it is interpreted as a (partial) Lagrangian mean position.} 


We now formulate two distinct strategies for the computation of the Lagrangian mean $\BARL f(\bx,\tau)$.

\subsection{Strategy 1}

Our first strategy  parallels that of \cite{GBLApaper} and consists in solving a PDE for $\tilde f (\bx,t)$, evaluating the result at $t=\tau + T$ to obtain $\TILDL{f}(\bx,\tau)$, then deducing  $\BARL{f}(\bx,\tau)$ by applying a suitable re-mapping. 
To derive the PDE for $\tilde f (\bx,t)$ we take the time derivative of \eqref{ftilde} at fixed label $\ba$ and use the chain rule and \eqref{flowmap} to find
\begin{align}
\partial_t \tilde f(\bphi(\ba,t),t) + \bu( \bphi(\ba,t),t) \bcdot \bnabla \tilde f(\bphi(\ba,t),t) = \frac{1}{t - \tau}  \left( f(\bphi(\ba,t),t) - \tilde f(\bphi(\ba,t),t) \right),
\end{align} 
Here and in what follows, the gradient $\bnabla$ is taken with respect to the first argument of $\tilde f$. Replacing $\bphi(\ba,t)$ by $\bx$ as independent variable yields the sought PDE,
\beq
\partial_t \tilde f(\bx,t) + \bu(\bx,t) \bcdot \bnabla \tilde f(\bx,t) = \frac{f(\bx,t) - \tilde f(\bx,t)}{t - \tau},
\label{ftildeevol}
\eeq
which can be integrated in from $\tau$ to $t = \tau + T$ to find the total mean $\TILDL{f}(\bx, \tau) = \tilde f(\bx,\tau + T;\tau) $. This is a forced advection equation in which the forcing can be interpreted as a time-dependent relaxation of $\tilde f$ to $f$.
In a bounded domain, the solution of \eqref{ftildeevol} requires no boundary conditions since the differentiation $\bu(\bx,t) \bcdot \bnabla$ is along the boundary. The initial condition is that  $\tilde f(\bx,\tau) = f(\bx,\tau)$ so that the right-hand side is finite.

\begin{figure}
\begin{center}
\begin{tikzpicture}[scale=0.8,
    thick,
    dot/.style = {
      draw,
      fill = white,
      circle,
      inner sep = 0pt,
      minimum size = 4pt
    }
  ]
  \coordinate (O) at (0,0);
  \coordinate (A) at (0.8,1);
  \coordinate (B) at (7,1); 
  \coordinate (C) at (6,-1);
  
  \draw[gray, thick, fill=gray!10] (O) ellipse (1.8 and 1.8);
  \draw[gray, thick, fill=gray!10] (6,0) ellipse (2 and 2.5);
  \node at (0,-2.4) {label space};
  \node at (6,-3) {physical space $\subseteq \mathbb{R}^n$};
  \draw[shorten >= 2pt,-{Stealth[scale=1]},thick,blue] (A) to[out=50,in=150] node[above] {$\bm{\varphi}$} (B); 
  \node[anchor=north,yshift=-0.2em,blue] at (A) {$\bm{a}$};
  \node[anchor=north,yshift=-0.2em,purple] at (C) {$\bar{\bm{\varphi}}(\bm{a},t;\tau)$};
  \node[anchor=south,yshift=0.2em] at (B) {$\bm{x}$}; 
  \draw[shorten >= 2pt,-{Stealth[scale=1]},purple] (A) to[out=10,in=150] node[above] {$\bar{\bm{\varphi}}$} (C);
  \draw[shorten >= 2pt,-{Stealth[scale=1]},thick,teal] (C) to[out=100,in=200] node[above,xshift=-0.5em] {${\bm{\Xi}}$} (B);
  \draw[shorten >= 2pt,-{Stealth[scale=1]},thick,teal] (B) to[out=290,in=20] node[above,xshift=1.2em,yshift=-0.9em] {${\bm{\Xi}^{-1}}$} (C);
  \draw[thick, blue, fill=blue!20] (A) circle (.8mm);
  \draw[thick, black, fill=black!20] (B) circle (.8mm);
  \draw[thick, purple, fill=purple!20] (C) circle (.8mm);
 \end{tikzpicture}
\caption{Flow map $\bphi$, Lagrangian (partial) mean map $\bar \bphi$, lift map $\bXi$ and its inverse $\bXi^{-1}$.}
\label{fig:Xi}
\end{center}
\end{figure}
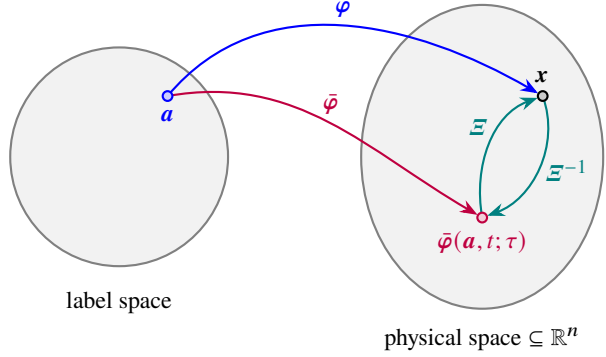

Computing $\TILDL f$ may be all that is needed for applications in which the spatial distribution of the Lagrangian mean is not important. For example, wave-averaged geostrophy -- the modified form of geostrophic balance expressed in terms of Lagrangian mean quantities -- 
can be validated by comparing Lagrangian mean velocity and pressure gradient at the same position (equivalently, the same label) regardless of where the position is located in physical space \citep{kafiabad2021wave,GBLApaper}.

However, in many other applications, it is necessary to compute the (generalised) Lagrangian mean field $\BARL f(\bx,t)$ for specified Lagrangian mean positions $\bx$.  This (generalised) Lagrangian mean field can be deduced from  $\TILDL f(\bx,t)$ by  considering the \textit{lift map} $\bXi(\bx,t)$ which returns the position at time $t$ of the particle with $\bx$  as  partial mean position from $\bar{t}$ to $t$, that is,
\beq
\quad \bXi(\BAR \bphi(\ba, t),t) =  \bphi(\ba,t) \label{Xi-def}
\eeq
\citep{andrews1978exact,buhler2009waves}.
Its inverse  $\bXi^{-1}(\bx,t)$  returns the partial mean position of the particle that passes through $\bx$ at $t$:
\beq
	\bXi^{-1}( \bphi(\ba,t) ,t) = \BAR \bphi(\ba,t) = \frac{1}{t-\tau} \int_{\tau}^t \bphi(\ba,s) \, \d s, \label{Xi-1-def}
\eeq
where we used the definition \eqref{barphi} of the mean position. 
The map $\bXi$ and its inverse $\bXi^{-1}$ are depicted in the figure \ref{fig:Xi}.
The relation \eqref{tildebar} between $\TILDL f$ and $\BARL f$ can then be written in terms of $\bXi^{-1}$ as 
\beq
\TILDL{f}(\bx,\tau) = \BARL{f}(\bXi^{-1}(\bx,\tau+T),\tau). \label{remap-strat1}
\eeq

Now, comparing  \eqref{Xi-1-def} with \eqref{ftilde} makes it clear that the components of $\bXi^{-1}$ can be viewed as instances of functions $\tilde f$ with $f(\bx)=x_i$. Thus, we can rewrite \eqref{ftildeevol} for this special case to obtain
\beq
\partial_t \bXi^{-1}(\bx,t) + \bu(\bx,t) \bcdot \bnabla \bXi^{-1}(\bx,t) = \frac{\bx - \bXi^{-1}(\bx,t)}{t - \tau}.
\label{Xi-1evol}
\eeq
Alternatively, we can take the time derivative of \eqref{Xi-1-def} and replace $\bphi(\ba,t)$ by $\bx$ to arrive at \eqref{Xi-1evol}. 
Integrating \eqref{Xi-1evol} provides the means to effect the remapping \eqref{remap-strat1} between $\TILDL f$ and $\BARL f$. 

To recapitulate,  our first strategy consists in solving the PDEs \eqref{ftildeevol} and \eqref{Xi-1evol} from $\tau$ to $\tau + T$ to obtain $\TILDL{f}(\bx,\tau)=\tilde f(\bx,\tau + T; \tau)$ and $\bXi(\bx,\tau+T; \tau)$, then using
\eqref{remap-strat1} to compute $\BARL{f}$ by interpolation. The algorithm proposed by \citet{GBLApaper} turns out to be a particular discretisation of this strategy (see \S\ref{sec:strat1impl} below).


\subsection{Strategy 2}

Our second strategy bypasses the use of $\tilde f$ and is instead based on PDEs for $\BAR f$ and $\bXi$. 
To derive these we first note that taking the time derivative of \eqref{barphi} gives
\begin{equation}
	\partial_t \BAR \bphi(\ba,t) = \frac{\bphi(\ba,t)-{\BAR \bphi(\ba,t)}}{t-\tau} \label{dbarphi_dt} \eqcolon \BAR \bu (\BAR \bphi(\ba,t),t). 
\end{equation}
The second equality defines the auxiliary velocity field $\BAR \bu $ as the time derivative of the partial Lagrangian mean position. Using \eqref{Xi-def}, this velocity field can be written in terms of the lift map as
\beq
\BAR \bu(\bx,t) = \frac{\bXi(\bx,t) - \bx}{t-\tau},
\label{baru}
\eeq
where the dummy variable $\bx$ can be thought of as the partial mean position.
We emphasise that $\BAR \bu(\cdot,t)$, like $\bXi(\cdot,t)$, depends implicitly on $\tau$ and warn that it should not be interpreted 
as the partial mean of the Lagrangian velocity: as discussed in appendix \ref{app:lmv}, its value at the end of the averaging interval, for $t=\tau + T$, differs from the usual Lagrangian mean velocity, that is, the time derivative of $\BARL \bphi(\ba,\tau)$ with respect to $\tau$.

Now, differentiating \eqref{fbar} with respect to $t$ at fixed label $\ba$ and using \eqref{dbarphi_dt} leads to
\beq
\partial_t \BAR f(\BAR \bphi(\ba,t),t) + \BAR \bu (\BAR \bphi(\ba,t),t) \bcdot \bnabla \BAR f(\BAR \bphi(\ba,t),t) = \frac{f(\bphi(\ba,t),t) - \BAR f(\BAR \bphi(\ba,t),t)}{t-\tau}. 
\eeq
We obtain the desired PDE for $\BAR f(\bx,t)$ by using \eqref{Xi-def} and replacing $\BAR \bphi(\ba,t)$ by the independent variable $\bx$ to write
\beq
\partial_t \BAR f(\bx,t) + \BAR \bu(\bx,t) \bcdot \bnabla \BAR f(\bx,t) = \frac{f(\bXi(\bx,t),t)-\BAR f(\bx,t)}{t-\tau}.
\label{barfevol}
\eeq
This is a forced advection equation, analogous to the PDE \eqref{ftildeevol} governing $\tilde f(\bx,t)$. However, unlike \eqref{ftildeevol} it is not closed since it involves $\bXi(\bx,t)$, explicitly on the right-hand side and implicitly through $\bar \bu(\bx,t)$ on the left-hand side. It needs to be solved along an equation for  $\bXi(\bx,t)$. We derive this equation by taking the  time derivative of \eqref{Xi-def} at fixed $\ba$ and using \eqref{dbarphi_dt} and \eqref{flowmap} to obtain
\beq
\partial_t \bXi(\bar \bphi(\ba,t),t) + \bar \bu(\bar \bphi(\ba,t),t) \bcdot \bnabla \bXi(\bar \bphi(\ba,t),t) = \bu(\bphi(\ba,t),t).
\eeq
Hence, replacing $\bar \bphi(\ba,t)$ by $\bx$ and using \eqref{Xi-def},
\beq
\partial_t \bXi(\bx,t) +\bar \bu(\bx,t) \bcdot \bnabla \bXi(\bx,t) = \bu(\bXi(\bx,t),t).
\label{Xievol}
\eeq
Strategy 2 consists in solving \eqref{barfevol} and \eqref{Xievol}, with $\bar \bu(\bx,t)$ defined by \eqref{baru}, for $t \in (\tau,\tau + T)$, then deduce the Lagrangian mean of $f$ as $\BARL f(\bx,\tau) = \bar f (\bx,\tau+T; \tau)$. 

The initial conditions for \eqref{barfevol} and \eqref{Xievol} are that $\bar f(\bx,\tau)=f(\bx,\tau)$ and $\bXi(\bx,\tau)=\bx$. 
The boundary conditions are non-trivial: in bounded domains, $\bar f(\bx,t)$ and $\bXi(\bx,t)$ are defined on the image of the label space by the Lagrangian mean map $\bar \bphi$ (equivalently, the image of the fluid domain by $\bXi^{-1}$). Thus, the problem in principle involves a boundary moving with velocity $\bar \bu$ and can therefore be difficult to discretise. The common situation where the physical domain has boundaries that coincide with constant-coordinate surfaces (curves in two dimensions) is straightforward, however, because the component-wise definition of $\bar \bphi$ in \eqref{barLbphi} ensures that it maps such boundaries to themselves so  the domain remains fixed. The case of periodic domains is also straightforward.

\section{Numerical implementation}\label{sec:num_impl}

The set of equations for each strategy of the previous section can be discretised in a variety of ways. Here we focus on a pseudospectral discretisation which we apply to the computation of  Lagrangian means in a turbulent shallow-water flow interacting with a Poincar\'e wave. We make general remarks about the choice of strategy and  numerical discretisation but leave a more complete analysis of numerical error and convergence for future studies.

\subsection{Strategy 1} \label{sec:strat1impl}

To solve \eqref{ftildeevol} it is convenient to introduce $\tilde g(\bx,t) = (t - \tau) \tilde f(\bx,t)$, leading to
\beq
\left(\partial_t + \bu(\bx,t) \bcdot \bnabla \right) \tilde g(\bx,t) = f(\bx,t). \label{gtilde-evol}
\eeq
Integrating this equation over the averaging period then yields the  Lagrangian mean 
$\TILDL{f}(\bx, \tau) = \tilde g(\bx,\tau + T)/T$. For simple geometries, periodic in particular, standard pseudospectral methods provide efficient solvers for  \eqref{gtilde-evol} and, if the remapping to Lagrangian mean positions is desired,
\eqref{Xi-1evol}. This is particularly convenient if the fluid's governing equations are also solved pseudospectrally, because $f$ is then available on spectral grid points to evaluate the right-hand sides of \eqref{gtilde-evol} and \eqref{Xi-1evol}, and on physical grid points to evaluate the nonlinear terms. An alternative is a semi-Lagrangian discretisation, which leads to the algorithm of \cite{GBLApaper} as detailed in appendix \ref{app:GBLA}. 

\begin{figure}
    \centering
     \includegraphics[trim=1cm 0 0 0, width=\linewidth]{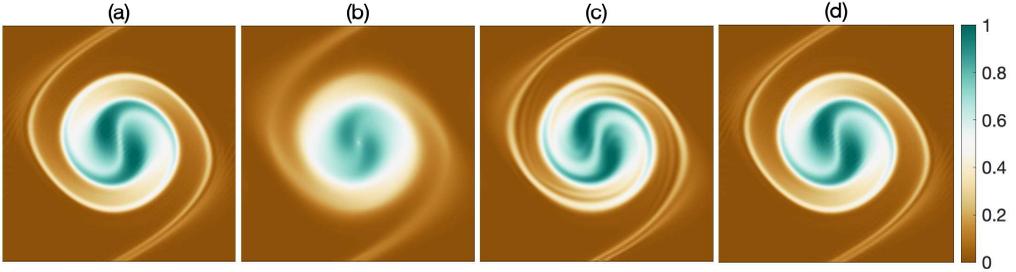}
\caption{Vorticity field and its Lagrangian mean for a two-dimensional incompressible inviscid flow: (a) instantaneous vorticity $\zeta$ at $t = 25$, and (b--d) $\tilL{\zeta}$ computed by time-integration of  \eqref{ftildeevol} from $\tau = 0$ to $T=25$ using (b) a semi-Lagrangian discretisation with linear interpolation, (c) semi-Lagrangian discretisation with cubic interpolation and (d) pseudospectral discretisation.}
\label{fig:vorticity2Dflow}
\end{figure}

To investigate the validity of numerical solutions of \eqref{ftildeevol} we consider a two-dimensional incompressible inviscid flow for which the vorticity, $\zeta$ say, is conserved materially. This implies that 
\begin{equation}
	\zeta(\bx,\tau + T) = \TILDL{\zeta}(\bx,\tau). \label{vorticity_itsmean}
\end{equation} 
Hence, we can calculate $\TILDL{\zeta}$ by integrating \eqref{ftildeevol} (or \eqref{gtilde-evol}) from $\tau$ to $\tau + T$ and compare it with the instantaneous vorticity at $\tau + T$ to study the accuracy of the computed Lagrangian mean. Note that this is simply a test for \eqref{ftildeevol} using the material conservation of $\zeta$ as opposed to an application of the Lagrangian mean. As mentioned earlier, \eqref{ftildeevol} should usually by solved in tandem with \eqref{Xi-1evol} to get a meaningful spatial distribution of Lagrangian mean quantities.

We perform a numerical simulation \HAKrev{of this two-dimensional flow without viscosity} in a doubly periodic, $\left[0,\ 2 \pi \right]^2$ domain, using a standard pseudospectral  discretisation \revv{and $2/3$ de-aliasing} with $128^2$ grid points. We start the simulation with the vorticity
\begin{equation}
 	\zeta(x,y,t=0) = \e^{-(x-\pi+0.1)^2 -(y-\pi+\pi/3)^2 }  +
 	                           \e^{-(x-\pi-0.1)^2 - (y-\pi-\pi/3)^2},  
\end{equation}
corresponding to two like-signed vortices which subsequently merge. We use Heun's method for the time integration of the governing vorticity equation and for \eqref{gtilde-evol}, with time step $ \Delta t = 0.005$. Figure \ref{fig:vorticity2Dflow} displays the instantaneous vorticity $\zeta$ at $t = 25$ and $\TILDL{\zeta}$ obtained for $\tau = 0$ and  $T = 25$. As expected from \eqref{vorticity_itsmean}, the Lagrangian mean $\TILDL{\zeta}$ matches the instantaneous vorticity $\zeta$. The pseudospectral solution for $\TILDL{\zeta}$, shown in panel (d), in particular, shows an excellent agreement with $\zeta$ in panel (a).  The results of the semi-Lagrangian  algorithm of \cite{GBLApaper} with, respectively, linear and cubic interpolations, are shown in panels (b) and (c) (see appendix \ref{app:GBLA}). These show a poorer agreement with panel (a), especially with the linear interpolation, because of an accumulation of interpolation errors. The computation reported in figure  \ref{fig:vorticity2Dflow} is however rather extreme in both the coarseness of the resolution  and the length of the averaging interval. 
We have confirmed that the three numerical solutions for $\TILDL{\zeta}$ converge to each other and to $\zeta$ as the spatial resolution increases or the length of the averaging interval decreases (\revv{the interested reader will find a demonstration of this convergence in the supplementary material}).  \revv{Since we solve the dynamical and Lagrangian-mean equations without explicit dissipation, both require small time steps. Our investigation for this particular example shows that the Lagrangian mean equations are less restrictive than the dynamical equations for the size of time step.}

The pseudospectral method  leads to the more the accurate results, but it is not as stable as its semi-Lagrangian counterpart and therefore requires smaller time steps. The difference arises because the implicit time integration and numerical smoothing due to interpolation that are inherent to semi-Lagrangian methods have a stabilising effect.

\subsection{Strategy 2}\label{sec:num_strategy2}

We now implement strategy 2 which uses the Lagrangian mean position $\bx$ as independent spatial variable. In the periodic domain we consider, there are no difficulties associated with moving boundaries and a pseudospectral discretisation is straightforward. It is convenient to rewrite the PDEs to be integrated,  \eqref{barfevol} and \eqref{Xievol}, in terms of the displacement map
\beq
\bxi(\bx,t) = \bXi(\bx,t) - \bx,
\eeq
since $\bxi$ is periodic, unlike $\bXi$. (This is the partial-mean analogue of the displacement introduced by \citet{andrews1978exact}.) As in strategy 1, it is also convenient to solve for $\bar g(\bx,t) = (t - \tau) \bar f(\bx,t)$ instead of $\bar f$. With these transformations, $\bar \bu = \bxi/(t-\tau)$ and \eqref{barfevol} and \eqref{Xievol} are rewritten as
\begin{subequations}
	\begin{align}
		\partial_t \bxi(\bx,t) +  \frac{\bxi(\bx,t)}{t-\tau} \bcdot \bnabla \bxi(\bx,t) &= \bu(\bx+\bxi(\bx,t),t) - \frac{\bxi(\bx,t)}{t-\tau} , \label{xi_evol}\\
	\partial_t \bar g(\bx,t) + \frac{\bxi(\bx,t)}{t-\tau} \bcdot \bnabla \bar g(\bx,t) &=  f(\bx+\bxi(\bx,t),t).  \label{barf_xi}
			\end{align}
	\label{setofxifbar}
\end{subequations}
These are the PDEs we solve numerically. When discretising in time, we found it beneficial for stability to  first update $\bxi$ using \eqref{xi_evol}, then use the updated $\bxi$ for the time integration of \eqref{barf_xi}. 

\HAKrev{Below we apply  strategy 2 to two examples of rotating shallow-water flows. We write the  governing equations in a non-dimensional form, using a characteristic length $L$, characteristic velocity $U$, time $L/U$ and mean height $H$ for scaling, leading to}
\begin{subequations}\label{sw_eqs_dimless}
     \begin{align}
     \frac{\partial \bu}{\partial t} + \bu \bcdot \bnabla \bu +  \frac{1}{\rm{Ro}} \hat{\boldsymbol{z}} \times \bu &= - \frac{1}{\rm{Fr}^2}\ \bnabla h + \frac{1}{\rm{Re}} \nabla^2 \bu, \label{sw_momentum_dimless}\\
       \frac{\partial h}{\partial t}  + \bnabla  \bcdot (h \bu) &=  0\ , \label{sw_mass_dimless}
     \end{align}
\end{subequations}
where we introduce the standard dimensionless numbers \citep[e.g.][]{vallis2017atmospheric}
\begin{equation}
	{\rm{Ro}} = \frac{U}{f L}\ , \quad {\rm{Fr}} = \frac{U}{\sqrt{g H}}\quad \textrm{and} \quad   {\rm{Re}} = \frac{U L}{\nu}.
\end{equation}
In the above,  $g$ is the gravitational acceleration, $\nu $ the kinematic viscosity, $f$ the Coriolis parameter and $\hat{\boldsymbol{z}}$ the vertical unit vector.

\subsubsection{Interaction of a turbulent geostrophic flow with a strong wave}\label{sec:SW_example1}

\begin{figure}
\centering
\includegraphics[width=.75\linewidth]{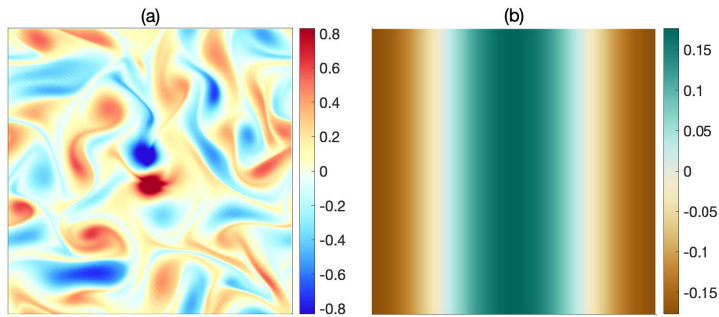}
\caption{Initial condition of the shallow-water simulation: (a)  vertical vorticity of the geostrophic  flow and (b) height field of the Poincar\'e wave.}
\label{fig:initial-cond}
\end{figure}

\HAKrev{In the first example,} we compute Lagrangian means in a simulation of a turbulent flow interacting with a Poincar\'e wave in a rotating shallow water. We initialise our simulation with a turbulent flow that is initially in geostrophic balance, with vorticity $\zeta = \partial_x v - \partial_x u$ shown in figure \ref{fig:initial-cond}(a), and superimpose a mode-1 Poincar\'e wave, with the height field shown in figure \ref{fig:initial-cond}(b). \HAKrev{The initial geostrophic flow is generated from the output of an incompressible two-dimensional Navier-Stokes simulation, which has reached a fully-developed turbulent state, and the height field that is in geostrophic balance with this vortical flow.} We use the root-mean square velocity of the geostrophic flow as the characteristic velocity $U$, and choose the length scale of first Fourier mode as characteristic length $L$. This makes the dimensionless doubly periodic domain $\left[0,\ 2 \pi \right]^2$ which we discretise with $256 \times 256$ grid points. The right-travelling mode-1 Poincar\'e  wave has the form 
\begin{equation}\label{initial_wave}
 u' =  a \cos (x-\omega t), \quad
 v' =  \frac{a}{\omega \rm{Ro}} \sin (x-\omega t), \quad
 h' =  \frac{a}{\omega} \cos (x-\omega t),
\end{equation}
with the intrinsic frequency $\omega = \left(\rm{Ro}^{-2}+\rm{Fr}^{-2}\right)^{1/2}$ and \JVrev{the velocity amplitude $a$ is a constant, taken as $a=-1.8$ in our simulation. This implies wave velocities that are almost twice as large as the geostrophic velocities. It is in this sense that we regard the wave as strong.} We set the dimensionless parameters  to 
\begin{equation}
	{\rm{Ro}} = 0.1 \ , \quad {\rm{Fr}} = 0.5 \quad \textrm{and} \quad  {\rm{Re}} = 3.84 \times 10^3,
\end{equation}
which results in $\omega = 10.2$.

We evaluate \eqref{initial_wave} at $t=0$ and add the wave fields to the geostrophic field to form the initial condition. We solve the dynamical equations \eqref{sw_eqs_dimless} in tandem with the Lagrangian mean equations \eqref{setofxifbar} over a single averaging time interval taken to be $T=2.2$, corresponding to approximately $3.6$ wave periods.
We use a \HAKrev{de-aliased} pseudospectral discretisation and a forward Euler integrator, with the time step of $1.25 \times 10^{-4}$ for \eqref{sw_eqs_dimless} and $2.5 \times 10^{-4}$ for \eqref{setofxifbar}. A bilinear interpolation is used to evaluate $\bu$ and $f$ at $\bx + \bxi(\bx,t)$ in the right-hand sides of \eqref{setofxifbar}. The link for the scripts and data used to produce the results of this section is provided in the ``Data availability statement" at the end of this paper.

\begin{figure}
\centering
\includegraphics[width=\linewidth]{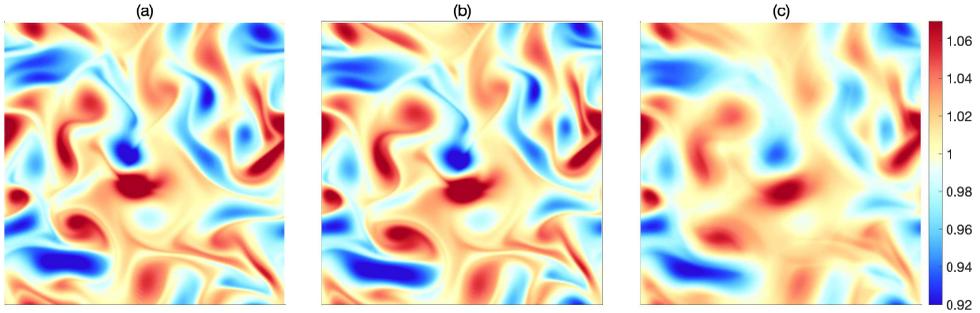}
\caption{Potential vorticity $q$ and its Lagrangian and Eulerian means in the shallow-water simulation: (a) instantaneous PV $q$ at $t = 1.1$, (b) Lagrangian mean $\barL{q}$, and (c) Eulerian mean PV. The mean fields are averaged from $\tau=0$ to $T=2.2$. All panels share the same colourbar.}
\label{fig:PVs}
\end{figure}

Figure \ref{fig:vorticities}, \JVrev{briefly discussed in \S\ref{sec:intro}},  displays the vorticity field $\zeta$ at $t=T/2$ (panel (a)) and its Lagrangian and Eulerian means (panels (b) and (c)). The strong wave dominates the instantaneous vorticity field. It is filtered successfully by time averaging. Clearly, the Lagrangian mean captures small-scale structures in the vorticity which are blurred by the Eulerian mean. This blurring is the result of the advection of the vorticity by the velocity field associated with the wave which causes the vorticity structures to oscillate with the wave period. By construction, the Lagrangian mean removes these oscillations, leading to a sharper definition of  the flow features.

It is interesting to examine the effect of Lagrangian and Eulerian averaging on the potential vorticity (PV)  \JVrev{$q=(1 + \rm{Ro}\ \zeta)/h$}. In the absence of dissipation ($\rm{Re} \to \infty$), this is a materially conserved field, meaning that $q(\bphi(\ba,t),t)=q_0(\ba)$, with $q_0$ determined by the initial condition. By definition \eqref{barLf}, the Lagrangian mean PV then satisfies $\barL q(\bar \bphi(\ba,t),t)=q_0(\ba)$. Thus both $q$ and $\barL q$  are (smooth) re-arrangements of the initial PV and hence re-arrangements of one another, specifically $\barL q(\bx,t)=q(\bXi(\bx,t),t)$. This imposes constraints such as the two fields sharing the same values for their local extrema. 
Because $\bar \bphi$ is not area-preserving, the distribution functions of $q$ and $\barL q$  (measuring the area of regions where the fields are below specified values) do not coincide.   Figure \ref{fig:PVs} shows $q$ at $t=T/2$ and $\barL q$ as well as the Eulerian mean PV. The Lagrangian mean PV $\barL q$ appears as a slight deformation of $q$, consistent with it being rearrangement by a map $\bXi$ that is close to the identity. In contrast, the Eulerian mean PV, which is not materially transported, shows blurred features, with in particular extrema that are substantially reduced compared with those of $q$ and $\barL q$. There is a strong argument that the study of wave--mean-flow interactions, in the shallow-water model and more broadly, would benefit from the systematic analysis of
Lagrangian mean fields such as the ones displayed in panels (b) of figures \ref{fig:vorticities} and \ref{fig:PVs}. 



\subsubsection{Moderate-Rossby-number flow initialised in geostrophic balance} 

\begin{figure}
\centering
\includegraphics[width=\linewidth]{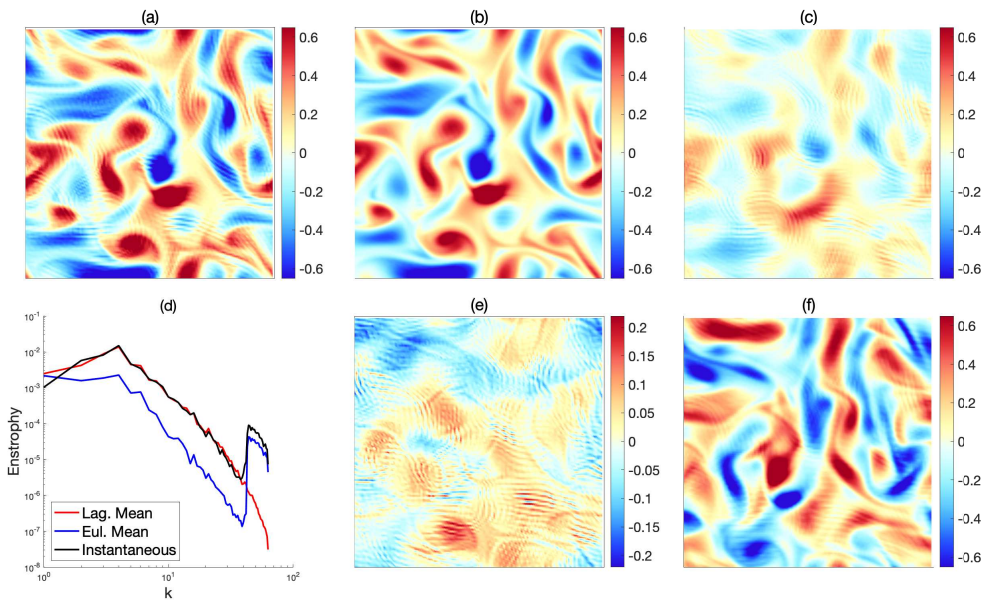}
\caption{(a) Instantaneous vorticity $\zeta(\bx,t)$ at $t=1.1$; (b) Lagrangian mean vorticity $\barL{\zeta}(\bx,\tau=0)$ corresponding to  averaging from $t=0$ to $T=2.2$. (c) Eulerian mean vorticity averaged from $t=0$ to $2.2$; (d) Enstrophy spectra of instantaneous velocity at $t=2.2$ and the Lagrangian and Eulerian mean fields; (e) Lagrangian perturbation field; (f) Eulerian perturbation field.}   
\label{fig:6panel_exmp2}
\end{figure}

\HAKrev{In the second example, we initialise the shallow-water simulation with the same geostrophic flow as in \S \ref{sec:SW_example1}  but without any waves. The set-up is identical, but with a resolution of $128^2$ and an increased Rossby number  $\rm{Ro} = 0.6$. As the flow evolves at this relatively high $\rm{Ro}$, small scale imbalance is generated as shown in the instantaneous vorticity of panel (a) in figure \ref{fig:6panel_exmp2} and the associated \revv{enstrophy} spectrum in panel (d). Panels (b) and (e) show the Lagrangian mean vorticity field and the corresponding perturbation field, respectively. The perturbation field is computed as $\BARL \zeta(\bx,\tau) - \zeta(\bx + \bxi(\bx,\tau+ T),\tau+T)$ with, in this case, $\tau=0$ and $T=2.2$. The Lagrangian average preserves the large-scale balanced flow so that the corresponding perturbation extracts the small-scale imbalance. This is evident in the enstrophy spectra of panel (d), where the instantaneous and Lagrangian mean spectra overlap at small wavenumbers and the jump in the tail of the instantaneous spectrum is removed in the Lagrangian mean spectrum. In contrast, the mean flow is weakened in the Eulerian average (panel (c)) because, at this moderate $\rm{Ro}$, the vortices are advected substantially during the averaging period. This is corroborated in the Eulerian mean enstrophy spectrum, which has lower values for small wavenumbers, and follows the jump of the instantaneous spectrum at large wavenumbers. Consequently, balanced flow features dominate the Eulerian perturbation field (panel (f)). The results of this figure are similar to those of the synthetic flow considered by \cite{shakespeare2021new}. }


\section{Discussion} \label{sec:discussion}

This paper presents a novel approach for the numerical computation of  Lagrangian means  which relies on solving PDEs rather than tracking particles. We propose two strategies, each leading to a separate set of PDEs. Both strategies are based on the derivation of equations governing the evolution of partial means with respect to $t$. These partial means are defined as averages over a subset $(\tau,t)$ of each averaging interval $(\tau, \tau +T)$ and yield the (total) means for $t=\tau + T$. 
Strategy 1 uses the position of particles at time $t$ as independent spatial variable. Hence, it requires a map from the positions at the final time $t=\tau + T$ to the Lagrangian mean positions to ultimately present the results in terms of the latter, as is standard in GLM theory. 
Strategy 2 directly computes the Lagrangian means in terms of Lagrangian mean positions. 

A natural question is  which of the two strategies should be preferred. There is no definitive answer: each strategy has pros and cons. If the spatial distribution of Lagrangian means does not matter in an application, it suffices to solve Eq.\ \eqref{ftildeevol} of strategy 1. When the spatial distribution is needed, strategy 1 requires the re-mapping \eqref{remap-strat1} which can be affected by clustering: the mean positions $\bXi^{-1}(\bx,\tau+T)$ obtained from \eqref{Xi-1evol} for $\bx$ on a regular grid may have a highly non-uniform distribution. This can lead to large numerical errors in the interpolation required to discretise \eqref{remap-strat1}. Strategy 2 circumvents this issue, as the generalised Lagrangian mean is computed directly on the desired spatial grid points. However, this advantage comes at the computational cost of evaluating more complicated right-hand sides in \eqref{barfevol} and \eqref{Xievol}, which require interpolation at each averaging time step.  Furthemore, strategy 2 leads to PDEs posed on a moving domain, unless the domain is periodic or has boundaries that correspond to  constant coordinates. 

As discussed in \cite{GBLApaper}, the full potential of our approach in saving memory and reducing computational cost is realised when the PDEs for the Lagrangian mean fields are solved on-the-fly, together with the dynamical model (as opposed to offline, using saved model outputs).  In this case, it is beneficial to solve the Lagrangian mean PDEs using a numerical scheme that closely matches that of the dynamical model, because the instantaneous values of $f$ and $\bu$ (required to solve the Lagrangian mean PDEs) are readily available at the same (physical or spectral) grid points. Moreover, the reasons that led to a particular choice of numerical discretisation for the dynamical equations -- such as the type of boundary conditions -- typically also apply to the Lagrangian mean PDEs.

In the main body of the paper, we restrict our attention to the Lagrangian averaging of a scalar function $f(\bx,t)$. The averaging of vectors, differential forms and more general tensors is however of interest in applications. For instance, the Lagrangian mean of the momentum 1-form $\bu \bcdot \d \bx$ (the integrand in Kelvin's circulation) and of the magnetic flux 2-form play crucial roles in the theory of wave--mean-flow interactions in fluid dynamics and MHD \citep{sowa72,andrews1978exact,holm2002lagrangian,gilbert2018geometric,gilbert2021geometric}. The derivations in \S\ref{sec:PDEs} generalise straightforwardly to tensors when the language of push-forwards, pull-backs and Lie derivatives is employed. We illustrate this in appendix \ref{app:tensor} by generalising Eq.\ \eqref{barfevol} of strategy 2 for the partial Lagrangian mean of $f(\bx,t)$ to a tensor field $\tau(\bx,t)$. 

We conclude by noting that practical averages such as the time average in \eqref{ave}--\eqref{barLf} do not satisfy exactly the axioms of the more abstract averages assumed in the development of GLM and similar theories. In particular, the basic requirement that averaging leaves mean quantities unchanged, that is, $\bar{\bar g} \not= \bar g$, fails for \eqref{ave},
though the difference is  small if there is a clear time-scale separation between mean flow and perturbation. Interpreting numerically computed Lagrangian mean fields in light of these theories will therefore require to understand how theoretical predictions are affected by the precise nature of the average.

\appendix

\section{Partial and total Lagrangian mean velocities} \label{app:lmv}

We show that the partial mean velocity $\bar \bu$ in \eqref{baru} 
is not related to the Lagrangian-mean velocity $\BARL \bu$ as might be expected naively: $\BARL \bu(\bx,\tau) \not= \BAR{\bu}(\bx,\tau + T;\tau)$, where we have reinstated the parameteric dependence on $\tau$ for clarity. 
The definition of $\BARL \bu$ itself is not without ambiguity: the most natural definition is as the derivative of the \textit{total Lagrangian mean map} with respect to the \textit{slow time}, i.e.\
\beq
\partial_{\tau} \BARL \bphi(\ba,\tau) = \BARL \bu(\BARL \bphi(\ba,\tau),\tau).
\label{ooo}
\eeq
Taking the derivative of the definition \eqref{barLbphi} of $\BARL \bphi$  gives the explicit form
\beq
\BARL \bu(\BARL \bphi(\ba,\tau),\tau) = \frac{1}{T} \left( \bphi(\ba, \tau + T) - \bphi(\ba, \tau) \right).
\label{as}
\eeq
In contrast, the partial mean velocity $\bar \bu$ is defined via a derivative with respect to the fast time $t$ as seen from the definition \eqref{dbarphi_dt} which takes the form  
\begin{equation}
	\partial_t \bar \bphi(\ba,t; \tau) \eqcolon \bar \bu(\bar \bphi(\ba,t),t; \tau).
\end{equation}
on reinstating the parametric dependence on $\tau$. 

The difference between $\bar \bu(\bx,\tau + T;\tau)$  and $\BARL \bu(\bx,\tau)$ should not come as a surprise since $\bar \bu(\bx,\tau + T;\tau)$ is constructed  independently for each value of $\tau$ and hence cannot capture the change of $\BARL \bphi(\cdot,\tau)$ measured by $\BARL \bu$. It can be made explicit by deducing from \eqref{as} that
\begin{align}
\BARL \bu(\bx,\tau) &= \frac{1}{T} \left( \bphi(\bar \bphi^{-1}(\bx,\tau + T;\tau), \tau + T) - \bphi(\bar \bphi^{-1}(\bx,\tau + T;\tau), \tau) \right) \nonumber \\ 
&= \frac{1}{T} \left( \bXi(\bx,t;\tau) - \bphi(\bar \bphi^{-1}(\bx,\tau + T;\tau), \tau) \right),
\end{align}
which differs from $\bar\bu(\bx,\tau+T;\tau)=\left( \bXi(\bx,\tau+T;\tau) - \bx \right)/T$ since $\bphi(\bar \bphi^{-1}(\bx,\tau + T;\tau), \tau) \not= \bx$. 

Note that in GLM,  $\BARL \bu$  is defined as the Lagrangian mean of the components of $\bu$ instead of via \eqref{ooo}. In Euclidean space the two definitions are equivalent since
\beq
\frac{1}{T} \int_{\tau}^{\tau + T} \bu (\bphi(\ba,s),s) \, \d s = \frac{1}{T} \int_{\tau}^{\tau + T} \partial_s \bphi(\ba,s) \, \d s = \frac{\bphi(\ba, \tau + T) - \bphi(\ba, \tau)}{T} ,
\eeq
using \eqref{flowmap}.
This equivalence does not hold for definitions of the Lagrangian mean flow that have been proposed as  geometric alternatives to GLM
valid on any manifold \citep{soward2010hybrid,gilbert2018geometric,vanneste2022stokes}.

\section{Semi-Lagrangian discretisation of strategy 1} \label{app:GBLA}



We show that the algorithm developed by \cite{GBLApaper} for the `grid-based calculation of Lagrangian average' is equivalent to a semi-Lagrangian discretisation of \eqref{ftildeevol} and \eqref{Xi-1evol}. We denote the grid points by $\bX_i$, with $i$ a multi-index in dimension more than one, and the set of all grid points by $\{ \bX_{\alpha} \}$. The semi-Lagrangian discretisation \eqref{ftildeevol} 
gives the value $\tilde{f}( \bX_i ,t_{n+1})$ of $\tilde f$ at each grid point at $t_{n+1}$, the end  of a time step, in terms of the set of values $\tilde{f}(\{ \bX_{\alpha} \},t_{n})$ at $t_n$, the start of the time step. It is based on a semi-Lagrangian discretisation of the equivalent PDE \eqref{gtilde-evol}, namely 
\begin{equation}
	\tilde{g}(\bX_i ,t_{n+1}) = \tilde{g}(\bY_i ,t_{n}) + (t_{n+1}-t_n) f(\bX_i ,t_{n+1}),  \label{sdiscrete_gtilde}
\end{equation}
where we use a backward Euler step for the time stepping. Here $\bY_i$ denotes the position at $t_n$ of the particle that passes through $\bX_i $ at $t_{n+1}$; it can  be evaluated using
\begin{equation}
	\bY_i = \bX_i - (t_{n+1}-t_n) \bu(\bX_i, t_{n+1}),
\end{equation}
which follows from approximating the derivative in \eqref{flowmap} by a backward finite difference. Rewriting \eqref{sdiscrete_gtilde} in terms of $\tilde{f}$ and rearranging we obtain
\begin{equation}
	\tilde{f}(\bX_i ,t_{n+1}) =  \frac{(t_{n}-\tau) \tilde{f}(\bY_i ,t_{n}) + (t_{n+1}-t_n) f(\bX_i ,t_{n+1})}{t_{n+1}-\tau},  \label{sdiscrete_ftilde}
\end{equation}
which is the same as Eq.\ (2.6) in \cite{GBLApaper} when the last term is replaced by their (2.7). Using the trapezoidal rule instead of backward Euler leads to  
\begin{equation}
	\tilde{f}(\bX_i ,t_{n+1}) =  \frac{(t_{n}-\tau) \tilde{f}(\bY_i ,t_{n}) + (t_{n+1}-t_n) \left[ f(\bY_i ,t_{n})+f(\bX_i ,t_{n+1}) \right] /2}{t_{n+1}-\tau},  \label{sdiscrete_gtilde_trapez}
\end{equation}
which is Eq.\ (2.8) of \cite{GBLApaper} and can provide more accurate results. Note that the evaluation of $\tilde{f}(\bY_i ,t_{n})$ requires an interpolation since $\bY_i$ is not on the grid. 

If we consider the particle position as a special case of $f$, \eqref{sdiscrete_gtilde_trapez} turns into
 \begin{equation}
	{\bXi^{-1}}(\bX_i ,t_{n+1}) =   \frac{(t_{n}-\tau) \bXi^{-1}(\bY_i ,t_{n}) + (t_{n+1}-t_n) (\bX_i+\bY_i)/2 }{t_{n+1}-\tau},  \label{sdiscrete_Xi-1}
\end{equation}
which parallels Eq.\ (2.9) in  \cite{GBLApaper}.  
The complete algorithm iterates \eqref{sdiscrete_ftilde}--\eqref{sdiscrete_Xi-1} from $t_0=\tau$ to $t_{N+1}=\tau + T$ to obtain $\TILDL f(\bx,\tau)=\tilde f(\bx,\tau + T)$ and $\bXi^{-1}(\bx,t+T)$. In the final step, $\bXi^{-1}(\bx,t+T)$
is employed to apply the  remapping \eqref{remap-strat1} to calculate $\BARL{f}(\bx, \tau)$.

\section{Lagrangian mean of tensors} \label{app:tensor}

We show how Eq.\ \eqref{barfevol} for the Lagrangian mean of scalar functions generalises readily to tensors when phrased in the language of push-forwards, pull-backs and Lie derivatives \citep[e.g.][]{fran04}. The definition \eqref{fbar} of the partial Lagrangian mean generalises as
\beq
\bar{\bphi}^{{\scriptscriptstyle \mathrm{L}}*}_{\tau} \BARL \tau_{\tau} \coloneq \frac{1}{T} \int_{\tau}^{\tau + T}  \bphi_s^* \tau_s \, \d s.
\label{geomglm}
\eeq
Here, we make the time variable appear explicitly as a subscript and we use $ 
\bar{\bphi}^{{\scriptscriptstyle \mathrm{L}}*}_t$ and $\bphi_t^*$ to denote the pull-backs by, respectively, the Lagrangian mean map and flow map. Pull-backs are interpreted differently depending on the nature of $\tau_t$: in particular, for a scalar, $(\bphi_t^* f)(\bx)=f(\bphi_t(\bx))$, while  for a 1-form, $(\bphi_t^* (\bu \bcdot \d \bx))(\bx)=u_j(\bphi_t(\bx)) \partial_i \varphi_{tj}(\bx) \, \d x_i$. The definition \eqref{geomglm} is a natural, geometrically intrinsic definition of the Lagrangian mean because it ensures that the tensors $\tau_s$ at different times $s$ are transported to the same position in label space before the averaging is carried out \citep[see][]{gilbert2018geometric}. 
The partial mean of $\tau_t$ associated with \eqref{geomglm} reads
\beq
\bar \bphi_t^* \bar \tau_{t} \coloneq \frac{1}{t-\tau} \int_{\tau}^{t}  \bphi_s^* \tau_s \, \d s
\label{bartau}
\eeq
and is understood to depend parametrically on $\tau$.

Differentiating \eqref{bartau} with respect to $t$ and using that $\partial_t \bar \bphi_t^* \bar \tau_t = \bar \bphi_t^* \left(\partial_t + \lie_{\bar \bu} \right) \bar \tau_t$, where $\lie_{\bar \bu}$ is the Lie derivative, gives \citep{gilbert2018geometric}
\beq
\bar \bphi_t^* \left(\partial_t + \lie_{\bar \bu} \right) \bar \tau_t = \frac{-\bar \bphi_t^* \bar \tau_t + \bphi_t^* \bar \tau_t}{t-\tau}.
\eeq
Pushing forward by $\bar \bphi_{t*}$ reduces this to
\beq
\left(\partial_t + \lie_{\bar \bu} \right) \bar \tau_t = \frac{\bXi_{t}^* \tau_t - \tau_t }{t-\tau}
\label{bartauevol}
\eeq
on using that \eqref{Xi-def}, here in the form $\bXi_t = \bphi_t \circ \bar \bphi_t^{-1}$, implies that $\bXi_t^* = \bar \bphi_{t*} \bphi^*_t$. Eq.\ \eqref{bartauevol} generalises \eqref{barfevol} to tensors. Its coordinate form depends on the type of tensor because of the different form taken by $\lie_{\bar \bu}$ and $\bXi_t^*$. Note that Eqs.\ \eqref{Xi-1evol} and \eqref{Xievol} for $\bXi^{-1}$ and $\bXi$ do not have a tensorial nature: because of the Cartesian definition of the Lagrangian mean map \eqref{barLbphi}, $\bXi$ is simply a triple of scalar functions rather than a vector. See \citet{gilbert2018geometric} for geometrically intrinsic definitions of the Lagrangian mean map alternative to  \eqref{barLbphi}.



\backsection[Funding]{HAK and JV were supported by the UK Engineering \& Physical Sciences Research Council (grant EP/W007436/1). JV was also supported by the UK Natural Environment Research Council (grant NE/W002876/1).}

\backsection[Declaration of interests]{The authors report no conflict of interest.}

\backsection[Data availability statement]{The data and scripts used for the simulations of \S\ref{sec:num_strategy2} are  available at \url{https://github.com/turbulencelover/ComputingLagMeans}}

\backsection[Author ORCID]{H. A. Kafiabad, https://orcid.org/
0000-0002-8791-9217; J. Vanneste, https://orcid.org/0000-0002-0319-589X}


%

\bibliographystyle{jfm}
\bibliography{myGFD.bib}

\end{document}